# Andreev terahertz radiation generators

N.T. Bagraev [1], L.E. Klyachkin [1], S.A. Kukushkin [2], A.M. Malyarenko [1], A.V. Osipov [2], V.V. Romanov [3], N.I. Rul' [1,3], K.B. Taranets [1,3]

[1] Ioffe Institute, St. Petersburg, Russia
[2] Institute for Problems in Mechanical Engineering of the Russian Academy of Sciences, St. Petersburg, Russia
[3] Peter the Great Saint-Petersburg Polytechnic University, St. Petersburg, Russia

**Abstract.** The electrical, magnetic and optical properties of edge channels consisting of spin circuits that contain single carriers in nanostructures of silicon, silicon carbide and cadmium fluoride are investigated. It is demonstrated that due to the presence of chains of negative-U dipole centers at the boundaries of the spin circuits, the latter are Andreev molecules for generating terahertz radiation.

## 1. Introduction

The terahertz (THz) range of the electromagnetic spectrum, which extends into the frequency range of 0.1−10 THz, remained inaccessible to researchers for a long time, which resulted in the appearance of the term "Terahertz Gap" in the scientific literature. This range is between the middle and far infrared (IR) range and the microwave radio frequency range. It received its name, firstly, because the radiation in this region of the spectrum is not transmitted by the Earth's atmosphere, and secondly, because of the absence of reliable THz sources and receivers operating at room temperature.

Historically, THz radiation has been of practical interest in astronomy [1](cosmic microwave background radiation) and in molecular spectroscopy, for studying vibrational and rotational modes [2]. In addition, this type of radiation has recently found more and more use in various fields, such as biology [3], medicine (tomography, sequencing) [4], security [5], the implementation of a new generation of 6G wireless communications using various THz sources and receivers are studied [6]. The use of photoconductive antennas [7, 8], operating at room temperature, emitting n a broad spectral range, and quite simple and cheap to manufacture, deserves attention in the latter case. However, he issues of generation and reception of induced multimode terahertz radiation remain unresolved to date.

The absence of compact, tunable coherent radiation sources, as well as low-noise receivers that do not require cooling to liquid helium temperatures has been the main constraint for these industries until recently.

Modern sources of THz radiation are divided into three types: thermal, electronic, and phononic. Thermal sources of THz radiation make it possible to obtain a broadband spectrum without the possibility of frequency tuning. Electronic sources of THz radiation are divided into vacuum and solid-state sources. Their main disadvantages are: size and high cost (synchrotrons, gyrotrons, free electron lasers), significant limitations on the frequency of the emitted radiation (the upper limit is about 1 THz) and its restructuring (traveling reverse wave lamps, resonant tunnel diodes, etc.). The main photonic device for generating THz radiation, which has found its practical application in IR Fourier spectroscopy, is a quantum cascade laser, which requires temperatures of ~ 261 K [9]. Small frequency tuning range is another significant problem for their wider application in various applied industries in addition to cryogenic temperatures.

Special attention should be paid to high-temperature superconductors (HTSC), as they outperform most sources of THz radiation both in the range of available frequencies and in the degree of tuning, however, they require cooling to temperatures of ~ 15−70 K [10].

Low-noise bolometers cooled to liquid helium temperatures are most often used as THz radiation receivers. However, the development of nanotechnology in the field of semiconductors has made it possible to obtain compact solid-state devices operating at room temperature, capable of acting both as sources and receivers of THz radiation [9].

Obtaining of THz radiation from the edge channels of silicon nanostructures due to the presence of shells consisting of chains of centers with negative correlation energy is demonstrated in this paper.

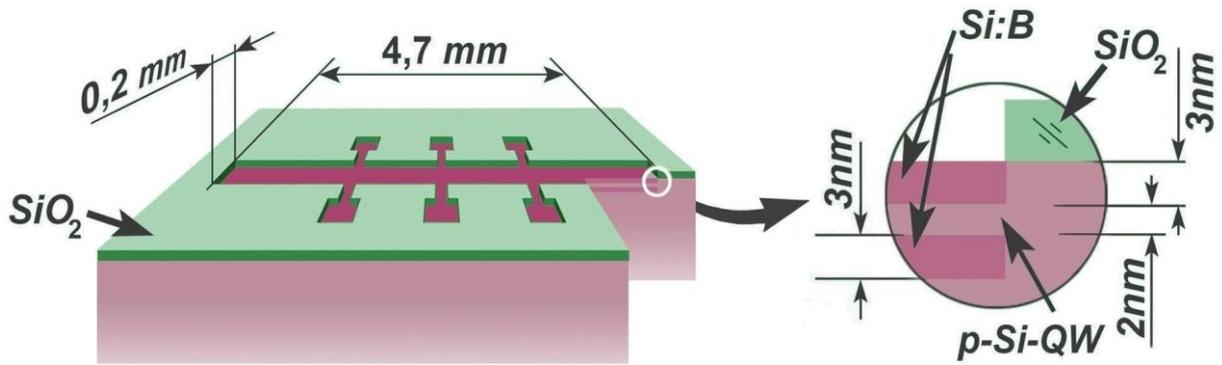

**Figure 1.** SNS design with typical dimensions.

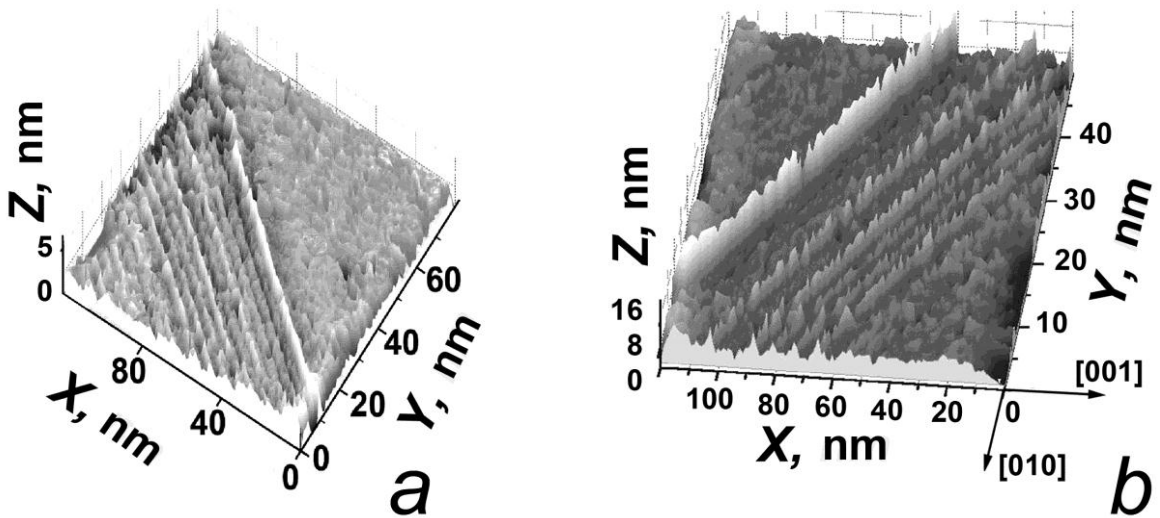

**Figure 2.** STM of the surface of a silicon nanosandwich (100) containing chains of boron dipole centers oriented along the axis (011).

## 2. Materials and methods

The nanostructure based on silicon, heavily doped with boron, used in this study has a number of characteristics that make it possible to observe broadband THz radiation with a high degree of coherence at room temperature. The silicon nanosandwich structure (SNS) of *p*-type of conductivity was obtained on silicon (100) *n*-type substrates in the process of preliminary oxidation and subsequent short-term gas-phase diffusion of boron [11]. It has been shown that boron atoms integrate into the crystal lattice, forming $\delta$-barriers (Figure 1) comprising chains containing boron dipole centers (Figure 2). Moreover, the dipole configuration of boron pairs is formed as a result of the so-called negative-U reaction: $2B^0 \rightarrow B^+ + B^-$, as a result f the reconstruction of boron atoms in the lattice sites, accompanied by the formation of trigonal pairs ($B^+ - B^-$) (Figure 3) [12]. The edge channel is a sequence of pixels containing single media (holes) that are captured on the negative-U of the boron chain [12]. It was found that tunneling of single holes through boron dipole centers leads to suppression of electron-electron interaction in the edge channels, and contributes to the observation of spin-dependent transport [13]. The studies of cyclotron resonance and spin-dependent recombination revealed an increase of the carrier relaxation time due to suppression of electron-electron interaction, which made it possible to observe such macroscopic quantum effects as the quantum Hall effect, the Shubnikov-de Haas and de Haas-Van Alphen, Aharonov-Bohm oscillations [12]. These measurements allowed determining the value of the two-dimensional density of holes, which was $p_{2D} = 3·10^{13}$ m$^{-2}$, which corresponds to the pixel length containing a single hole of 16.6 $\mu$m [12]. Moreover, each pixel (16.6 $\mu$m × 2 nm) can be considered as an Andreev molecule, inside which a single hole tunnels through boron dipole chains in opposite directions with an antiparallel spin orientation [12]. Such a pixel configuration can contribute to the realization of multiple Andreev reflection (MAR) within the Andreev molecule [14], which was initially demonstrated in studies of the magnetic and electrical characteristics of silicon nanosandwich [13].

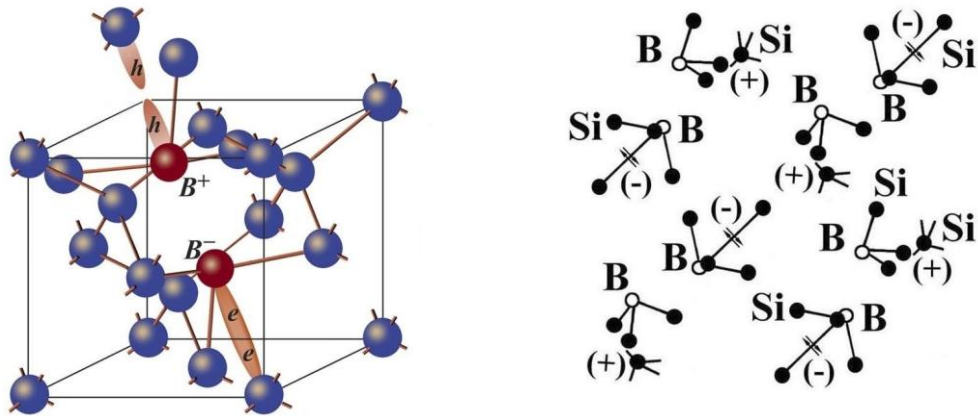

**Figure 3.** A boron dipole trigonal center ((B$^+$− B$^-$) with negative correlation energy and chains of boron dipole centers in $\delta$-barriers limiting an ultra-narrow silicon quantum well.

### 3. Results and discussion

The study of the field dependences of static magnetic susceptibility has shown that the change of spin and the movement of the hole in opposite directions can be controlled using an external magnetic field [13]. Moreover, the diamagnetic response of static susceptibility, detected in a weak magnetic field, indicated the presence of a correlation energy gap that controls the spin-dependent tunneling of a single hole through boron dipoles and, accordingly, contributes to the observation of macroscopic quantum effects at high temperatures [12]. The analysis of $T-B$-diagram of magnetic susceptibility shows that several correlation gaps appear in the SNS, characterized by a maximum value of negative magnetic susceptibility (Figure 4). $T-B$-diagram allows determining the values of the critical temperature and the critical magnetic field from the values corresponding to the maximum diamagnetic susceptibility, which was recorded when measuring the field dependences of static magnetic susceptibility at various temperatures [13]. The values of the critical temperature and magnetic field appear as a change of contrast on $T-B$ diagram. Thus, $T-B$-diagram demonstrates a sequence of energy gaps ($\Delta$) characterized by different values of critical temperature ($T_C$) and magnetic field ($2\Delta = 3.52 \, kT_C$): $2\Delta = 44$ meV, $T_C = 145$ K; $2\Delta = 33.4$ meV, $T_C = 110$ K; $2\Delta = 27.3$ meV, $T_C = 90$ K; $2\Delta = 22.8$ meV, $T_C = 75$ K; $2\Delta = 7.6$ meV, $T_C = 25$ K. Accordingly, the MAR current-voltage curve spectra were obtained for each correlation gap. This spectrum was obtained at a high temperature for the maximum value ($2\Delta = 44$ meV) [13] (Figure 5, *a* and 5, *b*). The above suggests that the sequence of correlation gaps occurs due to spin interference during the formation of single hole transport circuits in pixels of different widths relative to the boundary of the edge channel (Figure 6). The magnitude of the correlation gap decreases as its width increases. Thus, the pixel sequence across the edge channel can be considered as a system of Andreev molecules with different values of the correlation gap.

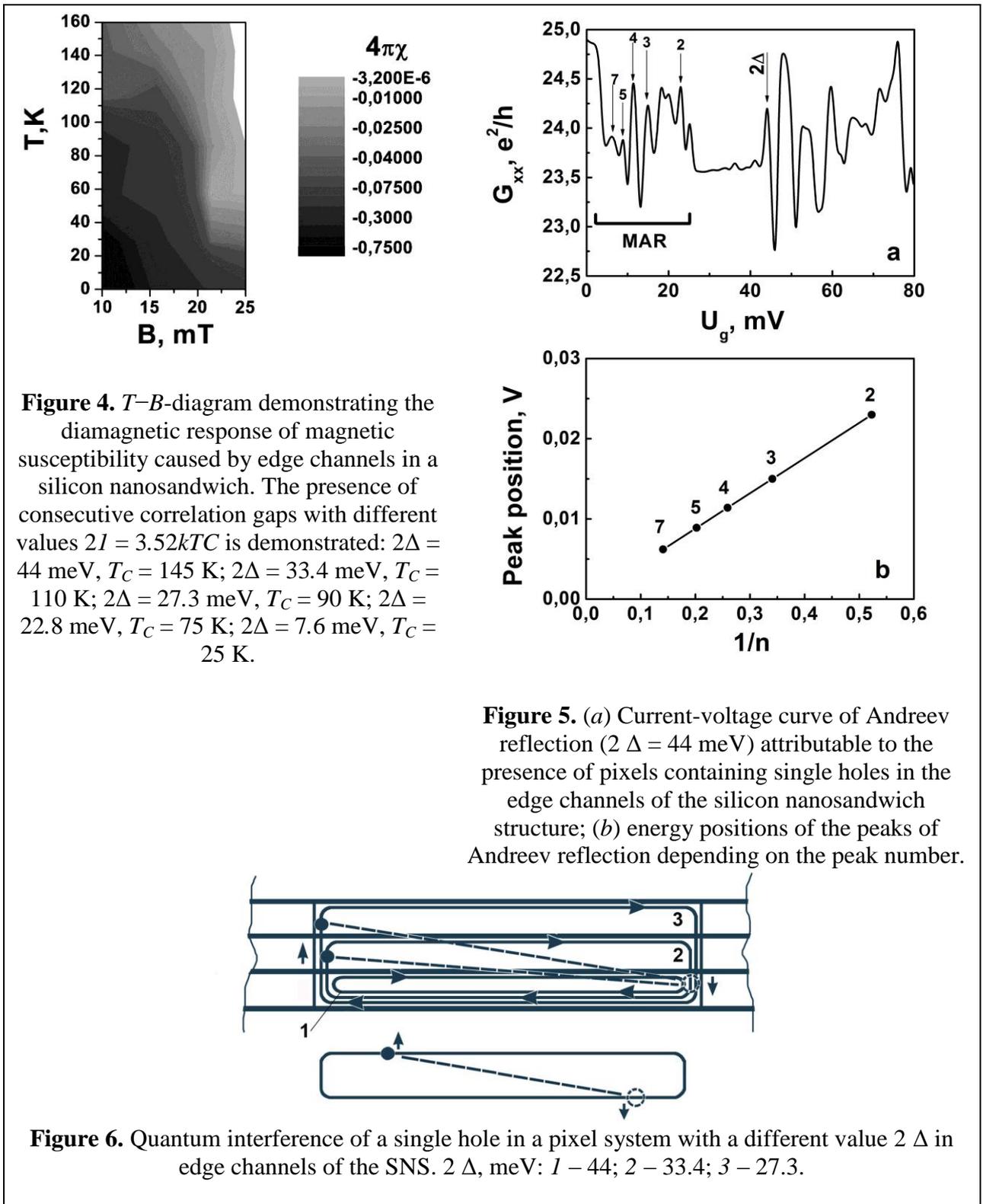

**Figure 4.** *T−B*-diagram demonstrating the diamagnetic response of magnetic susceptibility caused by edge channels in a silicon nanosandwich. The presence of consecutive correlation gaps with different values $2\mathit{1} = 3.52kTC$ is demonstrated: $2\Delta = 44$ meV, $T_C = 145$ K; $2\Delta = 33.4$ meV, $T_C = 110$ K; $2\Delta = 27.3$ meV, $T_C = 90$ K; $2\Delta = 22.8$ meV, $T_C = 75$ K; $2\Delta = 7.6$ meV, $T_C = 25$ K.

**Figure 5.** (*a*) Current-voltage curve of Andreev reflection ($2\Delta = 44$ meV) attributable to the presence of pixels containing single holes in the edge channels of the silicon nanosandwich structure; (*b*) energy positions of the peaks of Andreev reflection depending on the peak number.

**Figure 6.** Quantum interference of a single hole in a pixel system with a different value $2\Delta$ in edge channels of the SNS. $2\Delta$, meV: *1* − 44; *2* − 33.4; *3* − 27.3.

Since the tunneling of single holes along the edges of pixels in opposite directions corresponds to their antiparallel orientation, it is advisable to study the optical version of the MAR for identifying its spin-dependent component. The spectral characteristics of photo- and electroluminescence in the MAR energy range were studied by IR-Fourier spectroscopy using Bruker Vertex 70 spectrometers. The spectral dependences of electroluminescence for different gap widths are shown in Figure 7.

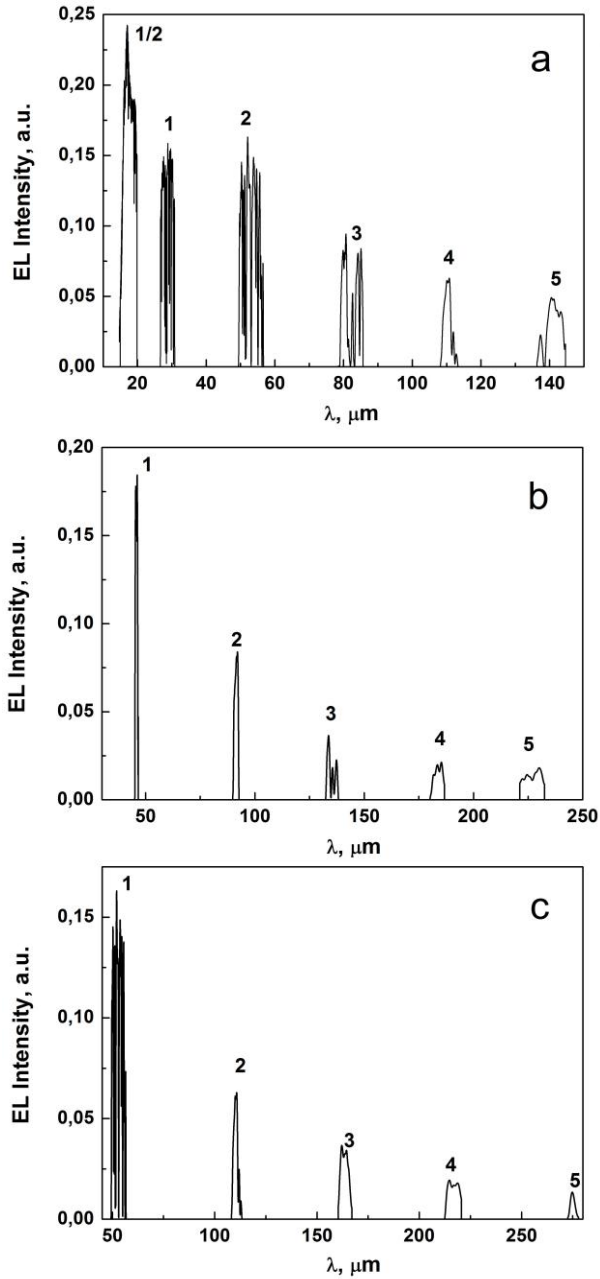

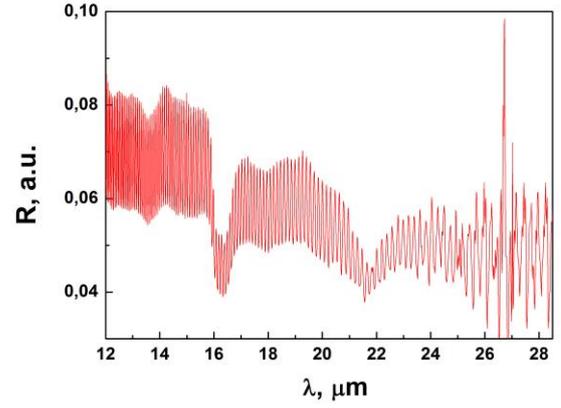

**Figure 8.** The transmission spectrum, which reveals the manifestation of both the local phonon mode, $\lambda = 16.4\ \mu m$, and the superconducting gap, $\lambda = 26.9\ \mu m$.

**Figure 7.** Induced radiation from the silicon nanosandwich structure attributable to MAR from the edge channels $T = 300K$ for the width of the correlation gap $2\Delta$, meV: $a – 44$, $b – 27.3$, $c – 7.6$.

The positions of the peaks in the electroluminescence spectra obtained for different values of the correlation gap energy are in good agreement with the energy transitions in MAR ($E = U_g/n$, where $n$ is the MAR peak number). The spectral line at $\lambda_{1/2}$ stands out for its amplitude, since it is consistent in this case with the characteristic dimensions of the pixel, which acts as a microresonator.

The mechanism of the optical version of MAR should take into account the spin-dependent component within the framework of the presented model, because, as mentioned above, the hole transport is accompanied by a spin flip in the spin circuit when the direction of movement changes. Thus, it can be assumed that the observed electroluminescence depends energetically on the magnitude of the spin-orbit interaction, i. e., the transition between chains of centers with negative correlation energy is accompanied by radiation with energy $h\nu \approx 2\Delta$ (for the energy of the largest gap, this value $2\Delta = 44$ meV corresponds to the magnitude of the spin-orbit interaction in the silicon valence band). The emitted quantum can be absorbed by a hole with the opposite spin orientation on a neighboring chain of centers with negative correlation energy and, thus, making a reverse transition, which is reflected in the corresponding transmission spectrum (Figure 8). Since there is a complete correspondence between the MAR emission spectra and the absorption spectra exhibited in case of the recording of the transmission spectrum, the sequence of these processes is responsible for the coherence of the transport of a single hole in the spin circuit. Thus, it can be assumed that the MAR inside single pixels in the edge channel is caused by the processes of spin-dependent transport [13]. The above was confirmed by studying the MAR spectral lines as a function of the gate voltage applied perpendicular to neighboring chains of centers with negative correlation energy. It is necessary to take into account the influence of the degree of polarization of a single hole during its transitions between chains in this case.

$$E(+z\ pol.) = \hbar^2 k_{x1}^2 / 2m^* - \alpha k_{x1}$$
$$E(-z\ pol.) = \hbar^2 k_{x2}^2 / 2m^* - \alpha k_{x2}$$
$$\Delta\theta = (k_{x1} - k_{x2})L = 2m^*\alpha L/\hbar^2$$
$$\Delta V_g \approx \frac{h^2 d^2 l}{3\pi^2 R m_{eff} \beta_{hh}}$$

The dependence of the spin-orbit interaction on the value of the transverse gate in the framework of this consideration conducted by Winkler [14] should lead to a shift of the peaks of spin-dependent electroluminescence to the region of higher or lower energies when the voltage sign on the transverse gate changes. Indeed, this effect is observed when studying the corresponding MAR spectral lines (Figure 9). Moreover, the magnitude of the spectral shift is consistent with the results of determination of electrical current-voltage curves of the spin transistor obtained on this silicon nanosandwich [15]. It should be noted that the spindependent nature of MAR indicates a possible role of the Majorana fermion in the transport of single holes due to the manifestation of the processes of transition with a spin flip between neighboring chains. In conclusion, it should be emphasized that within the framework of the experiment, multimode induced radiation from silicon nanostructures obtained within the framework of the Hall geometry was practically detected in a different frequency range, which is confirmed by the recorded dependence of the amplitude of spectral lines on the magnitude of the pulling current (Figure 10, *a* and 10, *b*). Similar results of the detection of spin-dependent induced radiation caused by MAR were obtained by studying nanostructures of silicon carbide and cadmium fluoride.

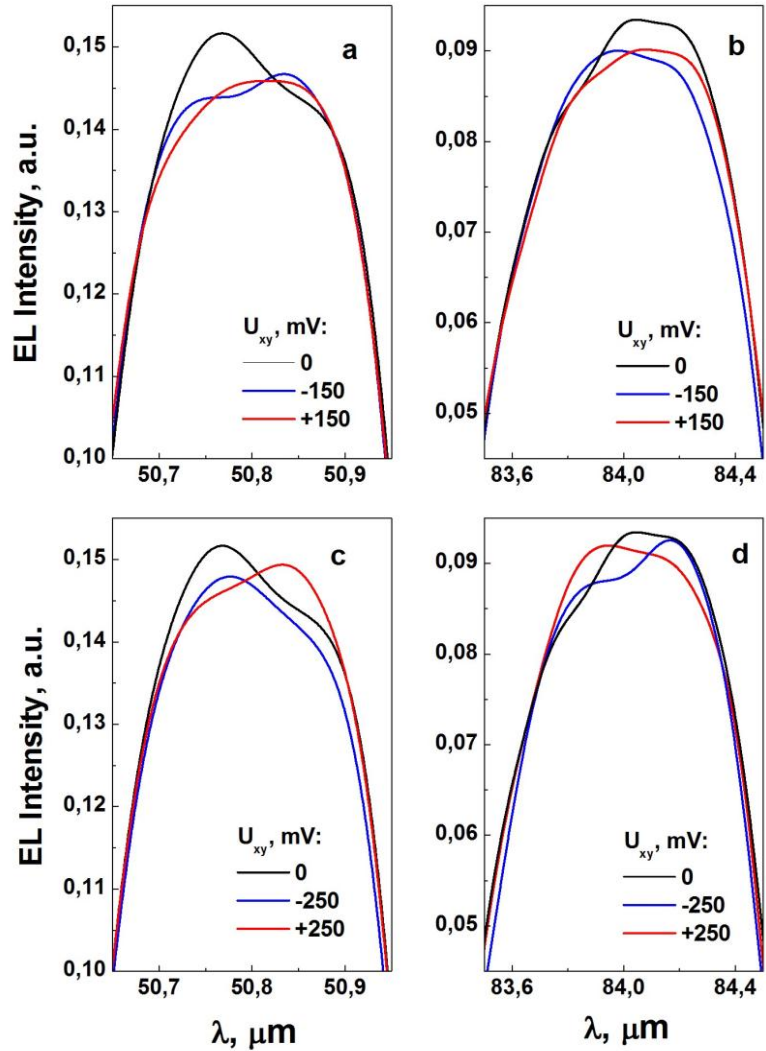

**Figure 9.** Spectral characteristic of the MAR line obtained at different values of the transverse gate voltage *a)* *1* — 0 mV, *2* — −150 mV, *3* — 150 mV; *b)* *1* — 0 mV, *2* — −150 mV, *3* — 150 mV; *c)* *1* — 0 mV, *2* — −250 mV, *3* — 250 mV; *d)* *1* — 0 mV, *2* — −250 mV, *3* — 250 mV.

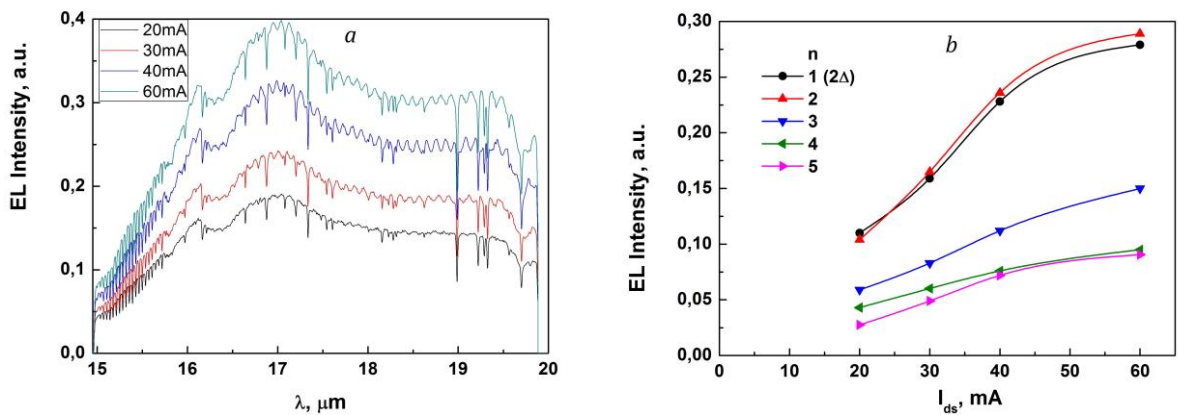

**Figure 10.** Induced radiation from silicon nanosandwich attributable to MAR (*1* = 22 meV) *a)* at different values of the pulling current *b)* dependence of the electroluminescence intensity on the magnitude of the pulling current (*n* — peak number).

## 4. Conclusion

It is shown that Andreev molecules in the edge channels of nanostructures can be used both as a source and receiver of THz radiation. The mechanism of emission and reception of THz radiation based on multiple Andreev reflection is presented.

**Funding**

The study was carried out within the research program planned by the Ioffe Institute of Physics and Technology, RAS.

**Conflict of interest**

The authors declare that they have no conflict of interest.